\documentclass[smallextended]{svjour3}     % onecolumn (second format)
\smartqed  % flush right qed marks, e.g. at end of proof
\usepackage{graphicx}
 \newcommand{\etal}{{\it et al.}}
\begin{document}

\title{The Abell cluster A586 and the detection of violation of the Equivalence Principle\thanks{\emph{Preprint:} DF/IST-4.2007%Grants or other notes
%about the article that should go on the front page should be
%placed here. General acknowledgments should be placed at the end of the article.
}
}
%\subtitle{Do you have a subtitle?\\ If so, write it here}

\titlerunning{A586 and Detection of Violation of Equivalence Principle}        % if too long for running head

\author{Orfeu Bertolami   \and
	Francisco. Gil Pedro   \and
        Morgan Le Delliou %etc.
}

\authorrunning{O. Bertolami,F. Gil Pedro \& M. Le Delliou} % if too long for running head

\institute{O. Bertolami and F. Gil Pedro \at
Departamento de F\'{\i}sica, Instituto Superior T\'{e}cnico \\
 Av. Rovisco Pais 1, 1049-001 Lisboa, Portugal\\
%              first address \\
%              Tel.: +123-45-678910\\
%              Fax: +123-45-678910\\
              \email{orfeu@cosmos.ist.utl.pt $\mathrm{and}$ fgpedro@fisica.ist.utl.pt%fauthor@example.com
}             \\
             \emph{O.B. Also at:} Centro de F\'{\i}sica dos Plasmas, IST.%of F. Author  %  if needed
           \and
           M. Le Delliou \at
              Instituto de Física Teórica UAM/CSIC, 
Facultad de Ciencias, C-XI,
Universidad Aut\'onoma de Madrid,
 Cantoblanco, 28049 Madrid SPAIN
              \email{morgan.ledelliou@uam.es}             \\
             \emph{Also at:} Centro de F\'{\i}sica Te\'orica e Computacional, Universidade de Lisboa, Av. Gama Pinto~2, 1649-003 Lisboa, Portugal, delliou@cii.fc.ul.pt
}

\date{Received: 18/12/08 / Accepted: date}
% The correct dates will be entered by the editor

\maketitle

\begin{abstract}
We discuss the current bounds on the Equivalence Principle, in particular
from structure formation and, reexamine in this context, the recent
claim on the evidence of the interaction between dark matter and dark
energy in the Abell Cluster A586 and the ensued violation of the Equivalence
Principle.
 \PACS{98.80.-k \and 98.80.Cq %\hfill \emph{Preprint:} DF/IST-4.2007%\and more
}
% \subclass{MSC code1 \and MSC code2 \and more}
\end{abstract}

\section{Introduction}
\label{intro}
The Equivalence Principle (EP) is a basic physical ingredient of general
relativity and shown to hold to great accuracy in composition dependent
free fall experiments involving known forms of matter. However, recent
discovery of empirical evidence of the existence of dark matter (DM)
in the cluster of galaxies $1E0657-56$, the so-called {}``bullet
cluster'' \cite{Clowe06}, and in a ring-like DM structure in the
core of the galaxy cluster $Cl0024+17$ \cite{Jee07}, together with
proposals on coupled dark energy (DE) have opened up the possibility
of having dark sector interactions which could alter the free fall
of dark matter (DM) and DE as compared to baryons.

In a recent paper \cite{Berto07}, we have argued that the study of
the virial equilibrium of the cluster Abell A586 indicates DE-DM interaction
and that this induces a detectable violation of the EP. This prompted
a discussion on the connection with existing related work on the free
fall in the presence of DM \cite{Stubbs93,Farrar06}%\cite{Stubbs93,KesdenKamion06,Farrar06}  
as well as on the phenomenological bounds on the interaction of DE
and DM arising from cosmology \cite{Guo07}.

In this paper we examine the consistency of the results of Bertolami \etal~\cite{Berto07}
with the existing literature. Our discussion is organized as follows:
in section \ref{sec:Two-kinds-of} we briefly summarize the way EP
violations can be detected. We then compare our previous results with
homogeneous cosmology data in section \ref{sec:EP-violation-impact}
and express them in terms of free fall in structure formation in section
\ref{sec:EP-violation,-free}. Section \ref{sec:Discussion-and-Conclusions}
contains our conclusions.

\section{{An EP Violation Typology}}

\label{sec:Two-kinds-of}One can identify two qualitatively distinct
forms of violation of the EP: i) A more classic one, where one observes
a differential acceleration on the free fall of baryons in a gravitational
field, the so-called fifth force type experiments and ii) A dark sector
violation, where one compares the free fall of baryons with the one
of DM or DE in a cosmological setting.

\subsection{{Baryonic composition dependent violation of the EP}}

In the fifth force type experiments one is concerned with the detection
of the difference on the free fall acceleration of baryonic matter
with different composition. A putative difference would be due to
a composition dependent new interaction. Current bounds on such a
differential acceleration imply that it should be smaller than $0.3\times10^{-13}$ \cite{Adelberger}. 
Alternatively, one could consider a differential
gravitational acceleration for baryonic matter and for DM or DE (DM
and DE if these two entities are interacting). The situation with
DM-baryons composition dependent interaction was examined in Stubbs~\cite{Stubbs93} 
and the bound $3.7\times10^{-3}$ for the differential
acceleration of galactic DM on different baryonic test masses was
obtained. The latest bound however is $\left(-4\pm7\right)\times10^{-5}$ \cite{Adelberger}. 
For sure, existing laboratory tests of the EP
are irrelevant in what concerns its validity on the dark sector. For
a general discussion of the EP validity for known particles and antiparticles,
see, for instance, Bertolami \etal~\cite{BPT06}.

\subsection{{Dark sector violation of the EP} }

The fact that recent observations suggest that about 96\% of the content
of the universe is composed by dark material of yet unknown nature
opens up the possibility of considering models where the free fall
of dark sector particles is different from the one of the baryonic
sector. Since the EP relies on clustering of mass, a violation of
the EP would be detected by comparing the baryonic free fall with
the clustering component of the dark sector: DM. That entails the
two possibilities for such modification: either the self-gravity of
DM gets modified in the \textit{particular} fashion suggested in Kesden and Kamionkowski~\cite{KesdenKamion06}, 
or this modification is induced by an additional
interaction between clustered and unclustered components of the dark
sector, DM and DE, respectively \cite{Berto07}. Contrary to the previous
case, detection cannot rely on free fall experiments of different
baryonic materials, but must involve baryon and DM falling onto DM \cite{KesdenKamion06,Berto07}.

Such configuration involves the formation of large scale structure,
at ranges where DM plays a leading role. In Bertolami \etal~\cite{Berto07},
we have used the gravitational relaxation in a galaxy cluster, A586,
and its peculiar type of statistical equipartition of energy to test
for the presence of modifications with respect to the dust behaviour
globally assigned to clustering matter. We have detected such modification
and identified it with the DE-DM interaction, which itself logically
entails a violation of the EP. We have proposed in Bertolami \etal~\cite{Berto07}
that such violation should be detectable in a generic bias present
at the homogeneous level. In section \textbf{\ref{sec:EP-violation,-free}},
we show that a particular model of DE-DM interaction yields a variation
of the gravitational coupling equivalent to the effects of the \textit{ad
hoc} model considered in Kesden and Kamionkowski~\cite{KesdenKamion06}. Such variation
is used there to separate the free fall of baryons (stars) onto the
DM filled galactic centre at a different rate than that of DM sub-haloes
of the Milky Way.

\section{{DE-DM Interaction and EP Violation}}

\label{sec:EP-violation-impact}
\begin{figure}%[pb]
%\begin{centering}\includegraphics[width=1\columnwidth,bb = 0 0 200 100, draft, type=eps]{GuoRescaled.eps}\end{centering}
\centerline{\includegraphics[width=1\columnwidth]{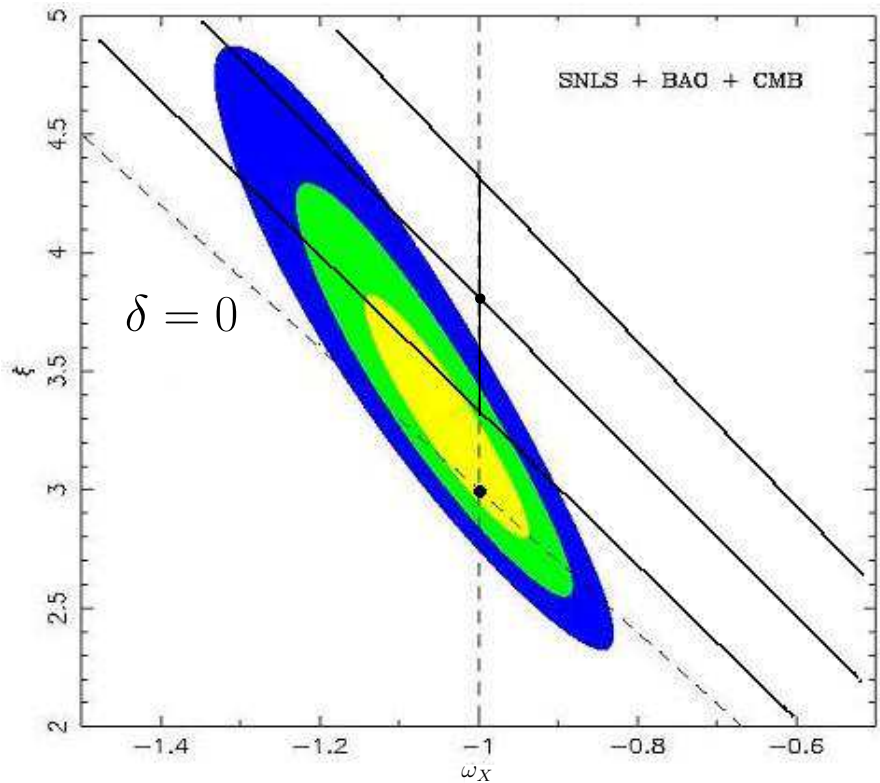}}
\vspace*{8pt}
\caption{Superimposition of the probability
contours for the interacting DE-DM model in the ($\omega_{X}$,$\xi$)
plane (denoted as ($\omega_{DE_{0}}$,$\eta$) in Bertolami \etal~\cite{Berto07}),
marginalized over $\Omega_{DE_{0}}$ in Guo \etal~\cite{Guo07} study of
CMB, SNIa and BAO (2.66$<\xi<$4.05 at 95\% C.L.) with the extended
results of Bertolami \etal~\cite{Berto07} based on the study of A586 cluster.
The $\xi=-3\omega_{X}$ dashed line corresponds to uncoupled models.
The result of Bertolami \etal~\cite{Berto07} is set at $\omega_{X}=-1$ and error
bars are shown. The thick lines corresponds to their extension to
varying $\omega_{X}$.\label{cap:Superimposition-of-the}}
\end{figure}

Models of interactions in the dark
sector have an impact on the EP since they modify the conservation
of DM, at least at the homogeneous level. Such modification entails
detectability on the evolution of the homogeneous geometry of the
universe, as shown for the phenomenologically consistent unified model
of DE and DM \cite{Bento4}, the generalized Chaplygin gas (GCG) model \cite{Bento02} 
or for the more generic interacting model \cite{Guo07,AmendolaRosenfeld07},
but also on the quality of the self-gravitating equilibrium reached
by large amounts of DM in clusters \cite{Berto07}. In Guo \etal~\cite{Guo07},
the consistency of two DE-DM interaction models, a constant coupling
model and a varying coupling model \cite{Amendola}, with respect
to the conjunction of data from Cosmic Microwave Background (CMB)
radiation shift parameter \cite{CMBshift}, SNLS observations of type
Ia Supernovae (SNIa) \cite{SNLS06} and the Baryon Acoustic Oscillation
(BAO) peaks in the SDSS \cite{BAOinSDSS}. The varying coupling model
was also considered, together with the GCG model, in the study of
the A586 cluster \cite{Berto07}, in which detection of the DE-DM
interaction was inferred from departure of the virial equilibrium.
In what concerns the varying coupling model, observational evidence
favors the detection of interaction in the region where both approaches
are compatible, namely for $\omega_{X}<-1$ (see Fig. \ref{cap:Superimposition-of-the}).%

The absence of detection in the homogeneous study \cite{Guo07} at
1 $\sigma$ level can be explained by the redshifts of CMB, SNIa and
BAO, which are much larger than the one of A586, while in the regime
of dominance of DE, the latter object is more affected. Should we
have access to more systems like A586, an error analysis would be
possible, adding the virial equilibrium as a Gaussian prior to obtain
a joint likelihood analysis with the homogeneous constraints. With
only one system to analyze, the signature shown on Fig. \ref{cap:Superimposition-of-the}
merely superimposes the results from Guo \etal~\cite{Guo07} with our study \cite{Berto07} 
with the respective errors. The overlap of contours
from both studies favors interacting phantom-type models ($\omega_{X}<-1$)
with energy transfer from DM to DE, as found in Bento \etal and Majerotto \etal~\cite{Bento04,AmendolaM}.

\section{{EP Violation, Free Fall and Structure Formation}}

\label{sec:EP-violation,-free}

In their simulation of a \textit{particular} modification of gravity
for DM, Kesden and Kamionkowski~\cite{KesdenKamion06} conclude that
the remnants of stars (baryonic matter) in the heading tidal streams
of the Sagittarius Milky Way satellite limits the variations induced
by the resulting variation of gravitational coupling, $G$, up to
10\%. Considering a model for the DE-DM interaction together with constant
mass DM particles, the effect on free fall gravitational force is
also to modify the coupling with time: the DM particle number density,
$n$, is diluted due to expansion and due to interaction with DE.
This interaction can be {\bf modeled} as an extra acceleration $a_{Int.}$
felt by DM particles on top of $a_{grav.}$, the gravity of the cluster.
Since this extra acceleration is only felt by DM particles, the violation
of the EP is explicit in this {\bf model}. Furthermore, using that the relative
change of total acceleration caused by this interaction is due to
the relative change in particle number density with respect to expansion
over a given cosmological time interval $\Delta t$: \begin{equation}
\frac{a_{Int.}}{a_{grav.}}\propto\frac{\left(\dot{n}_{DM}-\dot{n}_{DM}\Vert_{expansion}\right)\Delta t}{n_{DM}}~.\label{eq:RelIntDef}\end{equation}
 Since this is a cosmological effect, it is natural to consider $\Delta t\propto1/H$,
the Hubble time. The constant DM mass hypothesis leads to $\dot{\rho}_{DM}=\dot{n}_{DM}m_{DM}$,
which from the Bianchi identity, Eq. (1) from  Bertolami \etal~\cite{Berto07},
turns into \begin{equation}
\dot{n}_{DM}+3Hn_{DM}=\zeta Hn_{DM}~,\label{eq:DMcons}\end{equation}
 where \begin{equation}
\zeta=-\frac{(\eta+3\omega_{DE})\Omega_{DE_{0}}}{\Omega_{DE_{0}}+\Omega_{DM_{0}}a^{-\eta}}\:.\label{eq:DefZeta}\end{equation}
 Now, absence of interaction ($\zeta=0$), Eq.(\ref{eq:DMcons}) would
yield $\dot{n}_{DM}\Vert_{expansion}=-3Hn_{DM}$. Thus, introducing
a phenomenological constant $\delta$ that also absorbs the Hubble
time fraction, and using Eq. (\ref{eq:DMcons}), one obtains \begin{equation}
\frac{a_{Int.}}{a_{grav.}}=\delta\frac{\dot{n}_{DM}-\dot{n}_{DM}\Vert_{expansion}}{n_{DM}H}=\delta\zeta~.\label{eq:RelInt}\end{equation}
 Since this ratio depends only on the scale factor, $a$, the relative
change in acceleration is conveyed directly to the effective DM gravitational
coupling, $G_{DM}$: \begin{equation}
G_{DM}=G(1+\delta\zeta)~,\label{eq:Gdm}\end{equation}
 which allows for a straightforward comparison with the simulation
of Kesden and Kamionkowski~\cite{KesdenKamion06}. As this change is due to the interaction
with DE, which is significant only quite recently in the history of
the universe, $G_{DM}$ decreases as the strength of interaction becomes
more important. Notice that the interaction vanishes in the past and
then $G_{DM}(z)=G$. Given that the study of Kesden and Kamionkowski~\cite{KesdenKamion06}
sets from the observations a limit of about $\varepsilon=10\%$ as
the tolerance for such a change, we choose it to be the value for
$G_{DM}$ at present, therefore in the present model $\delta=\varepsilon/{\left|\zeta\right|}=0.163$,
assuming that $\omega_{DE_{0}}=-1$, $\Omega_{DE_{0}}=0.72$, $\Omega_{DM_{0}}=0.24$,
from which follows that%
\footnote{We point out that errors in Bertolami \etal~\cite{Berto07} were underestimated. The correct results 
are as follows: the virial ratio $\frac{\rho_{K}}{\rho_{W}}=-0.76\pm0.14\neq-0.50$,
the $\omega_{DE}=-1$ interacting pseudo-quintessence scaling $\eta=3.82_{-0.46}^{+0.50}\neq-3\omega_{DE}$
and the generalized Chaplygin gas power $\alpha=0.27_{-0.15}^{+0.17}\neq0$
. Conclusions remained unchanged.%
} 
$\eta=3.82_{-0.47}^{+0.50}$.
Notice that the negative sign, given by our observation of A586's virial ratio, arises from the fact that the detected energy transfer is from DM to DE, i.e. one has less DM and hence effectively a smaller effective gravitational coupling. For sure, one does not expect a negative overall coupling as the discussed effect is presumably small and ours is just a phenomenological
model of the complex process of interaction between concentrations
of DM, which are at the same time turning into DE.
It is interesting to remark that as the flow of energy density is
from DM to DE, then $n_{DM}$ is further decreased with respect to
expansion, which translates into a decrease on the coupling of DM
at present. Thus, one expects streams of stars to head the tidal arms
of absorbed satellites in this model.

Notice also that this calculation explicitly aims to translate the interaction
with DE into a phenomenological variation of gravity in clusters.
Of course, the value of $\varepsilon$ is chosen conservatively so
that the modification is compatible with the findings of Kesden and Kamionkowski~\cite{KesdenKamion06}.
However, it is fair to say that only a simulation with varying $G_{DM}$
would be fully conclusive. Actually, it is quite conceivable that
observations are consistent with even higher values of $\varepsilon$.

\begin{figure}%[pb]
\begin{centering}\includegraphics[width=1\columnwidth]{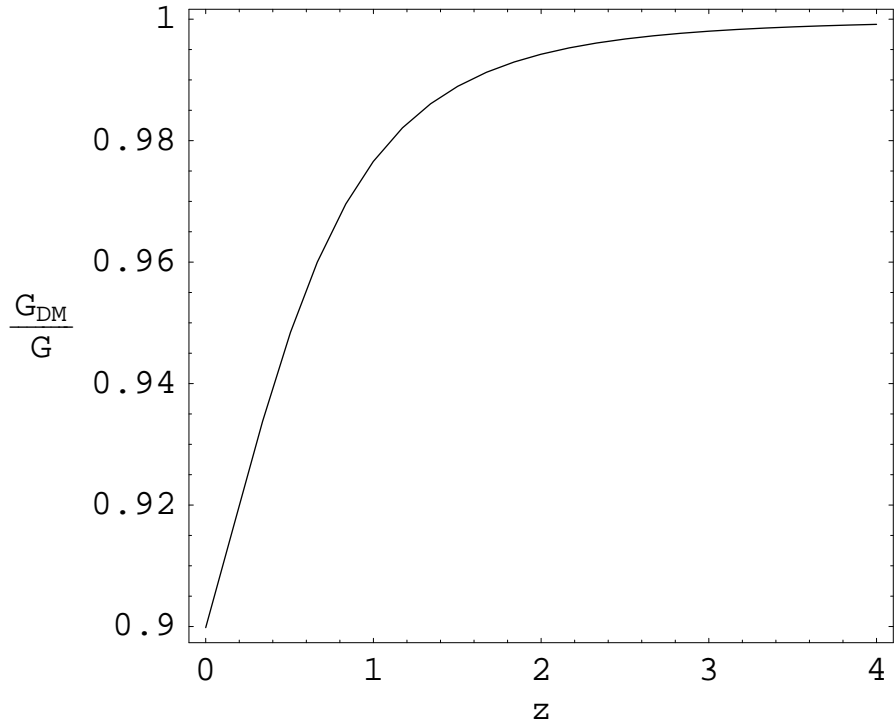}\par\end{centering}
\caption{\label{cap:Normalized-G-as} Evolution with redshift of the ratio
of the gravitational coupling for DM and baryons falling on a DM halo,
using the varying coupling model discussed in Bertolami \etal~\cite{Berto07},
as compared with the simulation of Kesden and Kamionkowski~\cite{KesdenKamion06}.}
\end{figure}

The behaviour of Eq. (\ref{eq:Gdm}) is shown in Fig. \ref{cap:Normalized-G-as}.

In the following, we discuss some of the salient features of the proposed
model. For the Milky Way satellite, the observed stream evolution
runs through at least 4 orbits that can be evaluated at 0.85 Gyr each.
This corresponds to redshifts between about $z=0.36$, where it started
with $G_{DM}/G=0.94$, and $z=0$, with a midpoint in evolution time
at $z=0.15$ where the ratio is $0.91$. For the A586, $z=0.17$,
and one obtains changes of $G_{DM}$ of about $8\%$ for the values
of Bertolami \etal~\cite{Berto07} (see above), as Eq. (\ref{eq:Gdm}) yields $G_{DM}/G=0.92$.
Notice that the observations of tidal streams from satellites of galaxies
thus far should reveal the kind of trailing studied in Kesden and Kamionkowski~\cite{KesdenKamion06}.
{[}Furthermore, the negative contribution to the effective gravitational
coupling can be attributed to the phantom-like nature of the interacting
DE model.]

It is relevant to point out that, in the context of the bullet cluster
($z=0.296$), an analysis of the velocity of the {}``bullet'' sub-cluster
indicates that a long range force with strength $G_{DM}/G=1.4-2.2$
is needed to account for the rather high observed velocities \cite{Farrar06},
even though a more elaborate analysis \cite{Springel07} suggests
that more detailed hydrodynamical models are required to reach a definite
conclusion. Our fit (which yields $G_{DM}/G=0.93$, for $z=0.296$
with the values from Bertolami \etal~\cite{Berto07}) does not account for such
a large deviation from the Newtonian regime. And clearly, a $\delta>0$
is not even qualitatively consistent with it, however, we stress that
the parameter $\delta$ reflects our ignorance on the %microphysics
correct model
of the interaction process and its details. In this context, it is
interesting to point out that if instead, $\delta$ were negative,
then our approach could, at least in principle, be consistent with
values required to accommodate the bullet cluster dynamics.

Finally, it is worth remarking that our results are consistent with
limits on the variation of the gravitational coupling with the energy
scale \cite{OBGBellido}.

\section{{Discussion and Conclusions}}

\label{sec:Discussion-and-Conclusions}In this paper we have examined
the compatibility of the existing studies on the interaction of dark
matter and dark energy and the results of Bertolami \etal~\cite{Berto07} concerning
the Abell A586 cluster. We have shown that there is no contradiction
with the available cosmological data, at least in what concerns the
varying coupling model of interaction between DE and DM. Furthermore,
we find that if one interprets the evidence of EP violation from the
Abell A586 as a change in the gravitational coupling, then our results
match those obtained in Kesden and Kamionkowski~\cite{KesdenKamion06} through a simulation
of the tidal stream of stars in the Sagittarius dwarf galaxy due to
DM. For sure, a more accurate comparison would involve considering
a simulation carried out in Kesden and Kamionkowski~\cite{KesdenKamion06} with the gravitational
coupling varying according to Eq. (\ref{eq:Gdm}). However, this is clearly
beyond the scope of this work.

It should be mentioned that an attractive new force or a stronger
gravitational coupling for DM might be an interesting solution to
many small discrepancies one encounters when attempting to fit data
with the $\Lambda$CDM model. These discrepancies include, the greater
than expected number of superclusters in the SDSS \cite{Einasto06},
the extreme low density of matter within the voids \cite{Farrar04},
the number of satellite galaxies which is an order of magnitude smaller
than predicted \cite{Moore99} among others. For sure, one can envisage
other fixes for these problems, such as, for instance, self-interacting
dark matter (see e.g. Bento \etal~\cite{Bento01} and references therein). Whatever
solution, one concludes that evidence for a complex and interacting
dark sector is mounting. It is then just logical to question whether
a basic physical ingredient such as the EP may hold for this sector
as well.

\begin{acknowledgements}
The work of MLeD is supported by CSIC (Spain) under the contract JAEDoc072,
with partial support from CICYT project FPA2006-05807, at the IFT,
Universidad Autonoma de Madrid, Spain, and was also supported by FCT
(Portugal) under the grant SFRH/BD/16630/2004, at the CFTC, Lisbon
University, Portugal. The work of
OB is partially supported by the FCT project POCI/FIS/56093/2004.
The authors would like to thank Christopher Stubbs, Marc Kamionkowski
and Rog\'{e}rio Rosenfeld for comments which prompt us to write this
letter. We are particularly thankful to Zong Guo for permission to
use a figure of Guo \etal~\cite{Guo07}.
%If you'd like to thank anyone, place your comments here
%and remove the percent signs.
\end{acknowledgements}

% BibTeX users please use one of
%\bibliographystyle{spbasic}      % basic style, author-year citations
%\bibliographystyle{spmpsci}      % mathematics and physical sciences
%\bibliographystyle{spphys}       % APS-like style for physics
%\bibliography{}   % name your BibTeX data base

% Non-BibTeX users please use

\end{document}